\documentclass[12pt]{article}

\usepackage{epsfig}  
\usepackage{axodraw}
\def\ii{\'\i}

\def\cao{\c c\~ao}

\def\ii{\'\i}

\def\cao{\c c\~ao}

\def\ftoday{{\sl {Le \number\day \space\ifcase\month 
\or janvier\or f\'evrier\or mars\or avril\or mai
\or juin\or juillet\or ao\^ut\or septembre\or octobre
\or novembre \or d\'ecembre\fi\space \number\year}}}    
\def\ptoday{{\sl {\number\day \space de\space \ifcase\month 
\or janeiro\or fevereiro\or mar{\c c}o\or abril\or maio
\or junho\or julho\or agosto\or setembro\or outubro
\or novembro \or dezembro\fi\space de\space \number\year}}}    
\def\gtoday{{\sl {Den \number\day. \ifcase\month 
\or Januar\or Februar\or M\"arz\or April\or Mai
\or Juni\or Juli\or August\or September\or Oktober
\or November \or Dezember\fi\space \number\year}}}    
\def\today{{\sl {\ifcase\month
\or January\or February\or March\or April\or May
\or June\or July\or August\or September\or October
\or November \or December\fi \space\number\day,\space 
                                            \number\year}}}
\newcommand{\journal}[4]{{\em #1~}#2\,(#3)\,#4}

\newcommand{\pr}{\journal {Phys. Rev.}}

\newcommand{\PRD}{{\em Phys. Rev. D}}

\newcommand{\AandA}{\journal{\em Astron. Astrophys.}}
 
\renewcommand{\d}{\delta}         
\newcommand{\e}{\varepsilon}

\newcommand{\la}{\lambda}        \newcommand{\LA}{\Lambda}
\newcommand{\m}{\mu}
\newcommand{\n}{\nu}
\newcommand{\om}{\omega}         \newcommand{\OM}{\Omega}

\newcommand{\f}{{\phi}}           

\newcommand{\MM}{{\cal M}}

\newcommand{\TT}{{\cal T}}

\newcommand{\es}{\\[3mm]}

\newcommand{\sla}{\raise.15ex\hbox{$/$}\kern -.57em} 
\newcommand{\Sla}{\raise.15ex\hbox{$/$}\kern -.70em}

\def\Lp{\displaystyle{\biggl(}}
\def\Rp{\displaystyle{\biggr)}}

\newcommand{\lp}{\left(}\newcommand{\rp}{\right)}

\newcommand{\complex}{{\kern .1em {\raise .47ex
\hbox {$\scriptscriptstyle |$}}
    \kern -.4em {\rm C}}}
\newcommand{\real}{{{\rm I} \kern -.19em {\rm R}}}
\newcommand{\rational}{{\kern .1em {\raise .47ex
\hbox{$\scripscriptstyle |$}}
    \kern -.35em {\rm Q}}}
\renewcommand{\natural}{{\vrule height 1.6ex width
.05em depth 0ex \kern -.35em {\rm N}}}

\newcommand{\dfud}[2]{{\displaystyle{\frac{\delta #1}{\delta #2}}}}
\newcommand{\dfrac}[2]{{\displaystyle{\frac{#1}{#2}}}}
   
\newcommand{\dint}{\displaystyle{\int}}
\newcommand{\eg}{{\em e.g.,\ }}
\newcommand{\Eg}{{\em E.g.,\ }}
\newcommand{\ie}{{{\em i.e.},\ }}

\newcommand{\twiddle}{\lower.9ex\rlap{$\kern -.1em\scriptstyle\sim$}}

\newcommand{\equ}[1]{(\ref{#1})}
\newcommand{\eq}{\begin{equation}}
\newcommand{\eqn}[1]{\label{#1}\end{equation}}
\newcommand{\eea}{\end{eqnarray}}
\newcommand{\eqa}{\begin{eqnarray}}
\newcommand{\eqan}[1]{\label{#1}\end{eqnarray}}
\newcommand{\ba}{\begin{array}}
\newcommand{\ea}{\end{array}}
\newcommand{\eqac}{\begin{equation}\begin{array}{rcl}}
\newcommand{\eqacn}[1]{\end{array}\label{#1}\end{equation}}

\newcommand{\bz}{\begin{enumerate}}
\newcommand{\ez}{\end{enumerate}}

\newcommand{\ADS}{(A)dS}

\begin{document}

\title{Dimensionally compactified Chern-Simon \\[2mm]
theory in 5D  
as a gravitation theory in 4D}

\author{Ivan Morales, Bruno Neves, Zui Oporto and 
Olivier Piguet\footnote{Talk presented by O.P. at the Conference  ``IWARA 2016: Quarks and Cosmos'', October 2016, Gramado, RS, Brazil.}
\\[4mm]
{\small Departamento de F\ii sica, Universidade Federal de 
Vi\c cosa (UFV)}\\
{\small  Vi\c cosa, MG, Brazil}
}

\date{}

\maketitle

\begin{center} 

\vspace{-5mm}

{\small\tt E-mails:
mblivan@gmail.com, bruno.lqg@gmail.com, azurnasirpal@gmail.com, opiguet@pq.cnpq.br \es
}
\end{center}

\begin{abstract}
We propose a gravitation theory in 4 dimensional space-time obtained by compacting  to 4 dimensions the five dimensional topological Chern-Simons theory with the gauge group SO(1,5) or SO(2,4) -- the de Sitter or anti-de Sitter  group of 5-dimensional space-time.  In the resulting theory, torsion, which is solution of the field equations as in any gravitation theory in the first order formalism, is not necessarily zero. However, a cosmological solution with zero torsion exists, which reproduces the Lambda-CDM cosmological solution of General Relativity. A realistic solution with spherical symmetry is also obtained.
\end{abstract}

{\small\it\noindent Keywords: Topological gravity; General Relativity; Cosmology.\\[2mm]
}
\section{Introduction}

Someone could ask: ``why are some people looking 
for alternatives to Einstein's General Relativity?''
After all, two theories describe very well all
natural phenomenons, at the  level considered nowadays 
as fundamental, at all scales which are technologically attainable:
from $10^{-26}$ s (\eg at LHC)  to 
$10^{17}$ s (the cosmological scale)
These theories are
General Relativity (GR), describing classical 
gravitation (a 101 years old theory!), and 
the Standard Model (SM), for electromagnetic and nuclear forces 
(which is a good 45 years old.).
No experiment or observation have contradicted any of both theories
... at least up to now.

Despite of this, the reader may perhaps find, 
in the following considerations, some justification for 
pursuing the search for alternatives -- apart the somewhat 
obvious observation
that, one being classical and the other quantum, something 
has to be done in order to reconcile these two theories.

\section{GR in the first order formalism}

Let us recall that in the original metric formalism GR 
is described by one tensor field 
$g_{\m\n}(x)$, the metric, and by two parameters: 
the Newton constant $G$  and the cosmological constant $\Lambda$. 
An equivalent description of GR is provided by the so-called
first order formalism, where field equations are of first order.
Let us consider a space-time of dimension $D$, with coordinates
$x=(x^\m,\,\ \m=0,\cdots,D-1)$
and metric $g_{\m\n}(x)$.
The objects are differential forms 
$f_\m(x)dx^\m$, $f_{\m\n}(x)dx^\m dx^\n$, etc.
The basic fields are

1. A pseudo-orthonormal basis for the 1-forms 
(``covariant vectors'') given by the ``$D$-bein'' forms
\eq\ba{l}
\phantom{with}\quad  e^I=e^I{}_\m(x) dx^\m\,,\quad I=0,\cdots,D-1\,,\es
\mbox{with}\quad 
e^I\cdot e^J := g^{\m\n}(x)e^I{}_\m(x)e^J{}_\n(x)=\eta^{IJ}\,,
\ea\eqn{D-bein}
where $\eta^{IJ} := \mbox{diag}(-1,1,\cdots,1)$
is the Minkowski metric, \ie the invariant quadratic form 
of the Lorentz group SO(1,$D$-1),
and $g^{\m\n}$ is the matrix inverse of $g_{\m\n}$
= $\eta_{IJ}e^I{}_\m(x)e^J{}_\n(x)$.

2. Lorentz connection 1-forms
\eq
\om^{IJ}=-\om^{JI}=\om^{IJ}_\m(x)dx^\m\,,
\eqn{Lorentz-connection}
defining the covariant derivatives. 
\Eg for a Lorentz vector $v^I(x)$:
$Dv^I = dv^I +\om^I{}_J\wedge v^J$,
with $d$ the exterior derivative,
such that $Dv^I$ be a Lorentz vector, \ie it transforms 
 under a local Lorentz transformation as
$(Dv^I)' = R^I{}_J Dv^J$, where $R^I{}_J$ is a SO(1,$D-1$) matrix.

Beyond invariance under the space-time diffeomorphisms 
(or general coordinate transformations), the theory is assumed to be
 invariant under the local Lorentz transformations 
 -- which constitute the gauge group of the theory.
 
The Einstein-Palatini action is the integral of a $D$-form, 
which, in the $D=4$ case, reads 
\eq
S_{\rm EP}[e,\om]=\dfrac{1}{64\pi G}\dint
\e_{IJKL} \lp e^I\wedge e^J\wedge R^{KL}
-\dfrac{\LA}{3} e^I\wedge e^J\wedge e^K\wedge e^L\rp,
\eqn{EP-action}
where $R^I{}_J:=d\om^I{}_J + \om^I{}_K\wedge\om^K{}_J$ is
the curvature 2-form and 
$\e_{IJKL}$ the completely antisymmetric Levi Civita tensor,
with $\e_{0123}=1$.
This action is equivalent to the 
more familiar Einstein-Hilbert action with cosmological
constant. Of course one has to add a contribution from matter fields $\f$:
$S_{\rm matter}[e,\om,\f]$.
The field equation obtained by varying the vielbein is the Einstein equation and the one obtained by varying the connection expresses the torsion 2-form $T^I:=De^I$ in terms of the matter fields, the result being a zero torsion if the matter action does not depend of the connection.

One may already note a reason for looking beyond GR: 
The cosmological term in the action \equ{EP-action} may
be put here or not, at will. Only observation~\cite{Planck-data} tells us that
it must be present\footnote{Or some ``dark matter'' artifact
giving results equivalent to the $\LA$CDM predictions.}, 
with a positive value 
$\LA\sim 10^{-35}$ s$^{-2}$ for its coefficient.
Moreover, the cosmological term 
is one of the many terms which may be put freely in the action,
using the metric defined above, 
such as higher powers of the Riemann tensor and torsion terms. This means a lack of necessity and uniqueness.


\section{(A)dS Chern-Simons theory for 5D gravity}

A more restrictive theory may be found in odd space-time 
dimension, as we shall see 
now\footnote{A very recent review on Chern-Simons gravity,
 with references to the original literature, 
 may be found in Ref.~\cite{hassaine-zanelli}. Kaluza-Klein
  compactification  in the more general framework of the 
  Lovelock-Cartan theory has been most recently discussed 
in  Ref~\cite{Castillo-etal}.}.
Let us consider a space-time dimension  $D=5$.

Doing as in dimension 4, we would get an action with terms 
$e\wedge  e\wedge  e \wedge  e \wedge  e$, 
$e\wedge  e\wedge  e\wedge  R$,  $e\wedge  R\wedge  R$,
written in symbolic form, each of them being 
multiplied by a free coefficient. 
Moreover, as in GR (see preceding Section), more Lorentz and 
diffeomorphism invariant terms may be added. 
However, let us  impose the larger local symmetry 
group SO(1,5) or SO(2,4), namely the  local $D=5$ de Sitter 
or anti-de Sitter group, denominated by 
(A)dS,  which contains the Lorentz group SO(1,4) as a subgroup.
The corresponding invariant action reads
\eq
S =  S_{\rm CS} + S_{\rm matter} 
 = \dfrac{1}{\kappa'} \int \OM_{\rm CS} + S_{\rm matter}\,.
\eqn{AdS-action}
$S_{\rm matter}$ is the part of the action describing matter 
and its interaction with the gauge field $A^{MN}$, 
whereas the integrand of the 
first term is the  \ADS\  Chern-Simons (CS) 5-form
\eq \ba{l}
\OM_{\rm CS} = \frac{1}{48}\e_{MNPQRS} \Lp
A^{MN}\wedge F^{PQ}\wedge F^{RS} 
-\frac12 A^{MN}\wedge (A\wedge A)^{PQ}\wedge F^{RS}\es
\phantom{\OM_{\rm CS} = \frac{1}{16}\e_{MNPQRS} \Lp}
+ \frac{1}{10} A^{MN}\wedge (A\wedge A)^{PQ}\wedge (A\wedge A)^{RS}
\Rp
\ea\eqn{CS-form}
where
 $A^{MN}=-A^{NM}$ is the (A)dS connection 1-form and
$ F^{MN}$ = $dA^{MN}+A^M{}_P\wedge A^{PN} $  is
its ``curvature''. The indices $M,\,N,\,\cdots$, run  from 0 to 5.
Recall the basic property of the CS form 
(written here according to the present group choice):
$d\OM_{\rm CS}=\e_{MNPQRS} F^{MN}\wedge F^{PQ}\wedge F^{RS}$.

We can interpret this  CS theory as a $D=5$ gravitation theory, writing a basis
 for the (A)dS Lie algebra in terms of the  10 generators of the Lorentz group, 
$M_{AB}=-M_{BA}$, and the  5 ``translation'' generators, $P_A$. Indices 
$A,\,B,\,\cdots$, run  from 0 to 4.
These generators obey the commutation rules
\eq\ba{l}
[M_{AB},M_{CD}]=  \eta_{BD} M_{AC} + \eta_{AC} M_{BD}
- \eta_{AD} M_{BC} - \eta_{BC} M_{AD}\,,\\[2mm]
[M_{AB},P_C] = \eta_{AC} P_B - \eta_{BC} P_A\,,
\quad { [P_A,P_B] = s\,M_{AB}}\,.
\ea\eqn{AdS-algebra}
$\eta_{AB}:=\mbox{diag}(-1,1,1,1,1)$   is the $D=5$ Minkowski metric,  
and the parameter $s$ takes the values $\pm1$ for the SO(1,5) de Sitter, respectively SO(2,4) anti-de Sitter group.

The expansion of the (A)dS connection in this basis is written as
\eq
A(x) = \frac12 \om^{AB}(x) M_{AB} + \dfrac{1}{ l}\, e^A(x) P_A\,,
\eqn{expand-connection}
where ${ l}$ is a parameter of dimension of a length,
necessary in order to match the dimension of 
the 5-bein forms $e^A$ (\,[length]$^1$\,)
to that of the Lorentz connection forms
$\om^{AB}$  (\,[length]$^0$\,).

With this notation, the gravitational part of the action \equ{AdS-action} reads
\eq\ba{l}
S_{\rm CS} = \dfrac{1}{8\kappa}\dint \e_{ABCDE} \Lp 
e^A\wedge R^{BC}\wedge R^{DE}
-\dfrac{2s}{3 l^2} e^A\wedge e^B\wedge e^C\wedge R^{DE}\\[2mm]
\phantom{S = \frac{\kappa'}{8}\int \e_{ABCDE} \Lp}
+ \dfrac{1}{5 l^4} e^A\wedge e^B\wedge e^C\wedge e^D\wedge e^E
\Rp\,,
\ea\eqn{CS-grav-action}
where $R^A{}_B=d\om^A{}_B+\om^A{}_C\wedge\om^C{}_B$ is
the Riemann curvature 2-form associated to the Lorentz connection $\om$.

This leads to the  field equations
\begin{eqnarray}
\dfud{S}{e^A} &=& \dfrac{1}{8\kappa} \e_{ABCDE}F^{BC}\wedge F^{DE}
+ \TT_A = 0 \,,\label{field-eq1}\\
\dfud{S}{\om^{AB}} &=& \dfrac{1}{2\kappa} \e_{ABCDE}T^C\wedge F^{DE}
+  \dfud{S_{\rm matter}}{\om^{AB}} =0 \,,\label{field-eq2}
\end{eqnarray}
where $T^A = D e^A := d e^A +\om^A{}_B\wedge e^B$ is the torsion 2-form, and
\eq
F^{AB}:=R^{AB} -\dfrac{s}{l^2} e^A\wedge e^B\,.
\eqn{FAB}
The energy-momentum 4-form $\TT_A$ := $ \d S_{\rm matter}/{\d e^A}$
is related to the energy-momentum components $\TT^A{}_B$ in 
the 5-bein frame by
\eq
\TT_A =\frac{1}{4!}\e_{BCDEF} \TT^B{}_A \,
e^C \wedge e^D \wedge e^E \wedge e^F.
\eqn{em-4-form}

 This is the theory.  Is it better than GR?

One aspect of it apparently makes it ``better''. 
Ba\~nados,  Garay and  Henneaux~\cite{Banados-etal} have shown that, in a class
of theories including the present CS theory, the 
invariance under the time diffeomorphisms follows from the invariance under 
the space diffeomorphisms and \ADS\ gauge invariance. 
This means that, in a canonical quantization~\cite{Dirac}
 of the theory (\eg in 
the framework of Loop Quantum Gravity~\cite{Rovellibook}), 
the constraint associated to the time diffeomorphisms 
-- which represents the main difficulty in the quantization program of 
GR~\cite{WdW,Rovellibook} -- 
is a mere consequence
of the constraints associated to the space diffeomorphisms and the \ADS\ gauge group.

A second interesting aspect, already at the level of the classical theory, is the fact that the
CS action \equ{CS-grav-action} is  uniquely defined by the invariances under 
diffeomorphisms and the gauge group \ADS, in the absence of any a priori given 
exterior metric.  That is, the gravitational part of the action depends on the 
unique dimensionless coupling constant $\kappa$ in \equ{AdS-action}
and on the dimensional scale $l$ which we had to introduce in
the expansion \equ{expand-connection}.

Third, the theory should at least reproduce the good results of GR. 
We are now going to discuss this  in the next Section.

\section{Solutions of physical interest}

\subsection{Solutions with constant curvature}

Obvious solutions are obtained as solutions of 
$F^{AB}=0$ (see def. \equ{FAB}).
In  terms of four dimensional spherical coordinates $r,\,\theta,\,\f,\,\chi$ and of 
a time coordinate $t$, they are given by the metric
\eq\ba{l}
ds^2=-(1- \frac{s}{l^2}r^2)dt^2 
+ \dfrac{1}{1- \frac{s}{l^2}r^2} dr^2\es
\phantom{ds^2=}
+ r^2 \lp d\theta^2+\sin^2(\theta)d\f^2
+\sin^2(\theta)\sin^2(\f)d\chi^2 \rp\,.
\ea\eqn{ADS5metric}
This is the { de Sitter ($s=1$) or anti-de Sitter ($s=-1$)
metric}, with the { cosmological constant} identified as
\eq
\LA=3\frac{s}{l^2}\,.
\eqn{cosmological-constant}
We already see, and this will be confirmed by more physical examples
in the  following subsections, that the presence of 
a cosmological constant is a prediction of the theory,  a consequence of 
the necessity of the dimensionfull parameter $l$ in 
\equ{expand-connection}.

\subsection{Kaluza-Klein compactification to $D=4$} 

We see our world as a  4-dimensional, de Sitter Universe at 
large scale. Hence let us choose the topology of 5-dimensional 
space-time such as it factorizes as $\MM_4\times S^1$, with 
$\MM_4$ ``our'' 4-dimensional space-time
and $S^1$ its compact subspace parametrized by the 4th space coordinate
$\chi$, $0\leq\chi\leq2\pi$.

All fields being periodic functions of $\chi$, a Fourier expansion yields
all the ``Kaluza-Klein modes''. We shall restrict in the following to the zero modes, \ie to solutions independent of $\chi$.

\subsection{Cosmological solutions}

The usual assumption of an isotropic and homogeneous 3-space yields a 
metric of the  FLRW type,
\eq
ds^2 = -dt^2 + \frac{a^2(t)}{1-k r^2} dr^2 + r^2 \lp d\theta^2 
+  \sin^2\theta\, d\f^2 \rp
+R^2\, d\chi^2\,,
\eqn{FLRW-metric}
(in the case of a flat 3-space) which may be obtained from 
the 5-bein
\eq
\lp e^A{}_\m \rp = \mbox{diag}\lp
1,\,\frac{a(t)}{\sqrt{1-k r^2}},\,a(t)r,\,a(t)r\sin\theta,\,R
\rp 
\eqn{FLRW-bein}
We are assuming  a zero torsion $T^A$
and a constant compactification scale $e^4{}_\chi=R$, 
hence $g_{\chi\chi}=R^2$ as in \equ{FLRW-metric}.

Matter will be represented by dust in physical 3-space, \ie 
a perfect fluid with zero pressure, described by the 
energy-momentum tensor (see Eq.~\equ{em-4-form})
\eq
\lp \TT^A{}_B\rp =\mbox{diag}\lp
-\frac{\rho(t)}{2\pi R},\,0,\,0,\,0,\,\la(t) \rp\,.
\eqn{dust-em-tensor}
The factor $1/(2\pi R)$, equal to the inverse of the 
compact subspace ``volume'', takes into account that 
 $\TT^{00}$ is the $D=5$ energy density, whereas 
 $\rho$ is taken here the $D=4$ energy density.

 The set of field equations \equ{field-eq2} is 
 automatically satisfied by 
 our assumption of zero torsion, whereas the set
  \equ{field-eq1} reduces 
 to three differential equations for $a(t)$, $\rho(t)$ 
 and $\la(t)$.
Two of them  are equivalent to the usual Friedmann 
equations~\cite{Friedmann-equations} for the scale 
parameter $a(t)$ and 
the energy density $\rho(t)$ (for zero pressure):
\eq\ba{l}
\ddot a/a -\LA/3 = -(4\pi G/3)\,\rho\,,   \quad
(\dot a/a)^2  -\LA/3 =(8\pi G/3)\,\rho\,.
\ea\eqn{Fried-eq}
The third equation gives the ``fifth dimension pressure''  
in terms of the scale:
\eq
\la(t)=\frac{1}{16\pi^2G R\LA a^3}
\lp \LA^2a^3-3\LA a \dot{a}^2 -3\LA a^2\ddot{a} 
+9\dot{a}^2\ddot{a}\rp\,.
\eqn{lambda(t)}
The following identifications of the 
coupling constant $\kappa$ and the length scale $l$ in 
terms of the Newton constant and the cosmological constant,
\eq
\kappa = (16 \pi^2/3)\,G R\LA\,,\quad l=\sqrt{3/\LA}\,,
\eqn{kappa-l}
have been used in order to arrive at the standard form 
\equ{Fried-eq} of the field equations.
With the big-bang boundary condition $a(0)=0$
this yields the known $\Lambda$CDM solution, 
the compactification  scale R remaining arbitrary. 
The parameter $\la(t)$, calculated from 
Eq.~\equ{lambda(t)}, turns out to obey an equation of state
\eq
\la(t) = -(2G/(\LA R))\, \rho^2(t)\,.
\eqn{eq-of-state}

\subsection{Solutions with spherical symmetry}

We again assume the torsion $T^A$ to be zero.
Assuming spherical symmetry in 3-space with coordinates 
$r,\,\theta,\,\f$,
we write the most general stationary metric as
\eq\ba{l}
ds^2 = -n^2(r) dt^2 +a^2(r) dr^2+ r^2 ( d\theta^2+\sin^2\theta \,d\f^2)\es
\phantom{ds^2 =} 
+ g_{\chi\chi}(r) d\chi^2 + 2 g_{r\chi}(r) dr \,d\chi \,.
\ea\eqn{spherical-metric}
We still assume that the $\chi$-part of the geometry is 
independent of 
the localization in 3-space, namely: 
\eq
g_{r\chi} = 0\,,\quad g_{\chi\chi}=R^2 \mbox{ independent of }r\,,
\eqn{local-indep}
and look for vacuum solutions. The ``vacuum'' is  defined 
by the 4-dimensional part of the energy-momentum tensor 
being vanishing, keeping a possibly non-zero 
``5th dimension pressure'' $\la$:
\eq
\lp \TT^A{}_B\rp =\mbox{diag}\lp
0,\,0,\,0,\,0,\,\la(t) \rp\,.
\eqn{sph-sym-em-tensor}
Using a 5-bein reproducing  the metric \equ{spherical-metric}, with
\equ{local-indep} taken into account:
\[
\lp e^A_\m\rp = 
\mbox{diag}\lp n(r),\,a(r),\,r,r \sin\theta,\,R\rp\,,
\]
we get the following solution of the field equations:
\eq\ba{l}
n^2(r) = 1-\dfrac{2\mu}{r}-\dfrac{\LA}{3}\,r^2\,, \quad
a^2(r) = \lp 1-\dfrac{2\mu}{r}-\dfrac{\LA}{3}\,r^2 \rp^{-1}\,, \quad
\lambda(r) = \dfrac{6\m^2}{\kappa} r^{-6}\,.
\ea\eqn{sol-sph-sym}
The metric is thus of the Schwarzschild-de Sitter type 
in 4D subspace-time, 
with Schwarzschild mass $\m$, which is an integration constant,
and with a fifth compactified dimension of arbitrary scale $R$.

The 5D energy-momentum tensor is not zero: there appears a 
non-vanishing
fifth dimension pressure $\lambda(r)$ decreasing as $1/r^6$.
This is the prize to be payed in order to have 
a non-trivial solution.

We finally note that the special case of a vanishing mass $\m$ represents
a $D=5$ space-time factorizing in a 4-dimensional \ADS\  space-time 
-- of constant curvature and zero torsion -- and a circle of 
arbitrary radius $R$. 
In this case, the ``fifth dimension pressure'' vanishes.

\section{Conclusions}

We have shown that the simplest possible model of gravity
based on a Chern-Simons theory for the gauge group \ADS\  together
 with Kaluza-Klein compactification can be considered as phenomenologically viable. An interesting feature is the necessity
 of a cosmological constant from purely algebraic ground: there is no possibility of replacing the \ADS\ group by the Poincar\'e group.
 
 More details will be published elsewhere~\cite{MNOP}.

\subsection*{Acknowledgments}

This work was partially funded by the
Funda\cao\ de Amparo \`a Pesquisa do Estado de Minas Gerais -- 
FAPEMIG, Brazil (O.P.),
the Conselho Nacional de Desenvolvimento Cient\'{\i}fico e
 Tecnol\'{o}gico -- CNPq, Brazil (I.M., Z.O.  and O.P.)
and the Coordena\cao\ de Aperfei\c coamento de Pessoal de N\ii vel Superior --
CAPES, Brazil (I.M. and B.N.).


\end{document}